\input harvmac
\newif\ifdraft\draftfalse
\newif\ifinter\interfalse
\ifdraft\draftmode\else\interfalse\fi
\def\journal#1&#2(#3){\unskip, \sl #1\ \bf #2 \rm(19#3) }
\def\andjournal#1&#2(#3){\sl #1~\bf #2 \rm (19#3) }

\def\ie{{\it i.e.}}
\def\eg{{\it e.g.}}

\def\etc{{\it etc}}
\def\p{\partial}
\def\ap{\alpha'}

\def\frac#1#2{{#1\over#2}}

\def\half{\frac12}

\def\inbar{\,\vrule height1.5ex width.4pt depth0pt}
\def\IC{\relax\hbox{$\inbar\kern-.3em{\rm C}$}}
\def\IR{\relax{\rm I\kern-.18em R}}
\def\IP{\relax{\rm I\kern-.18em P}}

%
%

\def\prl#1#2#3{Phys. Rev. Lett. {\bf #1} (#2) #3}

\catcode`\@=11
\def\slash#1{\mathord{\mathpalette\c@ncel{#1}}}
\overfullrule=0pt

\def\LL{{\cal L}}

\def\OO{{\cal O}}

\def\SS{{\cal S}}

\def\VV{{\cal V}}

\def\underrel#1\over#2{\mathrel{\mathop{\kern\z@#1}\limits_{#2}}}

\catcode`\@=12


%

\def\det{{\rm det}}

\def\det{{\rm det}}
\def\exp{{\rm exp}}


\def\[{[}
\def\]{]}

\def\comment#1{ }

%
\def\draftnote#1{\ifdraft{\baselineskip2ex
                 \vbox{\kern1em\hrule\hbox{\vrule\kern1em\vbox{\kern1ex
                 \noindent \underbar{NOTE}: #1
             \vskip1ex}\kern1em\vrule}\hrule}}\fi}
\def\internote#1{\ifinter{\baselineskip2ex
                 \vbox{\kern1em\hrule\hbox{\vrule\kern1em\vbox{\kern1ex
                 \noindent \underbar{Internal Note}: #1
             \vskip1ex}\kern1em\vrule}\hrule}}\fi}

%
%






               

%
%
\def\inbar{\hskip.3em\vrule height1.5ex width.4pt depth0pt}
\def\IC{\relax{\inbar\kern-.3em{\rm C}}}
\def\IN{\relax{\rm I\kern-.16em N}}
\def\IQ{\relax\hbox{$\inbar$\kern-.3em{\rm Q}}}
\def\IZ{\relax{\rm Z\kern-.8em Z}}
%
%

%

%
\rightline{EFI-2000-32}
\rightline{RUNHETC-2000-34}
\Title{
\rightline{hep-th/0009148}}
{\vbox{\centerline{ Some Exact Results on Tachyon Condensation}
\vskip 10pt
\centerline{in String Field Theory}}}
\bigskip
\centerline{David Kutasov,$^1$ Marcos Mari\~no,$^2$ and 
 Gregory Moore $^2$}
\bigskip
\centerline{$^1$ {\it Department of Physics, University of Chicago}}
\centerline{\it 5640 S. Ellis Av., Chicago, IL 60637, USA}
\centerline{kutasov@theory.uchicago.edu}
\bigskip
\centerline{$^2$ {\it Department of Physics, Rutgers University}}
\centerline{\it Piscataway, NJ 08855-0849, USA}
\centerline{marcosm, gmoore@physics.rutgers.edu}

\bigskip
\noindent
The study of open string tachyon condensation in string field
theory can be drastically simplified by making an appropriate
choice of coordinates on the space of string fields. We show
that a very natural coordinate system is suggested by the
connection between the worldsheet renormalization group  and
spacetime physics. In this system only one field, the tachyon,
condenses while all other fields have vanishing expectation
values. These coordinates are also well-suited to the study
of D-branes as solitons. We use them to show that the tension
of the D25-brane is cancelled by tachyon condensation and
compute exactly the profiles and tensions of lower dimensional
D-branes.
\vfill

\Date{September 19, 2000}

\newsec{Introduction}

\lref\sfttach{V. A. Kostelecky and S. Samuel, ``The static 
tachyon potential in the open bosonic string theory,'' Phys. Lett. {\bf 
B 207} (1988) 169; 
``On a nonperturbative vacuum for the open bosonic string,'' 
Nucl. Phys. {\bf B 336} (1990) 263.
A. Sen and B. Zwiebach, ``Tachyon condensation in string 
field theory,'' hep-th/9912249, JHEP {\bf 0003} (2000) 002. 
N. Moeller and W. Taylor, ``Level truncation and 
the tachyon in open bosonic string field theory,'' hep-th/0002237, 
Nucl. Phys. {\bf B 583} (2000) 105.}
\lref\lumps{J. A. Harvey and P. Kraus, ``D-branes as unstable 
lumps in bosonic open string theory,'' hep-th/0002117, JHEP {\bf 0004} 
(2000) 012. R. de Mello Koch, A. Jevicki, M. Mihailescu, and R. Tatar, 
``Lumps and $p$-branes in open string field theory,'' hep-th/0003031, 
Phys. Lett. {\bf B 482} (2000) 249. 
N. Moeller, A. Sen and B. Zwiebach, ``D-branes as tachyon 
lumps in string field theory,'' hep-th/0005036, JHEP {\bf 0008} 
(2000) 039.}
\lref\liwitten{K. Li and E. Witten, 
``Role of short distance behavior in off-shell open-string
field theory,'' hep-th/9303067,
Phys. Rev. {\bf D 48} (1993) 853.}
\lref\hkms{J.A. Harvey, S. Kachru, G. Moore, and E. Silverstein,
``Tension is dimension,''  hep-th/9909072, JHEP {\bf 0003} (2000) 001.}
\lref\elitzur{S. Elitzur, E. Rabinovici and G. Sarkissian,
``On least action D-branes,'' hep-th/9807161, Nucl. Phys. {\bf B 541} 
(1999) 731.}
\lref\gerasimov{A. Gerasimov and S. Shatashvili, ``On exact tachyon potential
in open string field theory,'' hep-th/0009103.}
\lref\witcs{E. Witten, ``Noncommutative geometry and string field 
theory,'' Nucl. Phys. {\bf B 268} (1986) 253.}
\lref\lcpp{A. LeClair, M.E. Peskin and C.R. Preitschopf, 
``String field theory on the conformal plane, 1. Kinematical 
principles,'' Nucl. Phys. {\bf B 317} (1989) 411.}

The problem of open string tachyon
condensation on unstable branes in bosonic and supersymmetric
string theory   is interesting, since it touches on important issues 
in string theory such
as background independence, off-shell physics, the symmetry
structure of the theory, and the role of closed strings.
In the context of string field theory (SFT)
the main approach to this problem has been through Witten's 
cubic, or Chern-Simons string field
theory  \witcs, and in the past year notable progress 
has been made (see \eg\ \refs{\sfttach,\lumps}).

On the other hand, the physics of tachyon condensation is well
understood from
the first quantized (worldsheet) point of view.
The endpoint of condensation is a state in which the brane
has completely ``disappeared.''  The process of condensation can
also produce lower dimensional unstable branes (or BPS brane
-- anti-brane pairs in the superstring) as intermediate states.

Reproducing these results in the SFT of \witcs\ is non-trivial.
The apparent simplicity of a cubic interaction vertex is deceptive
-- the condensation involves an infinite number of physical and
unphysical scalar fields of arbitarily high mass. Recent progress on
the problem involves a level truncation \sfttach,  which appears
to lead to very good agreement with the expected results
for some quantities, such as the vacuum energy after condensation.
At the same time, it is not clear why and when level
truncation works, and it is difficult to study the dynamics of the
non-trivial vacuum using this approach.

\lref\hkm{J. A. Harvey, D. Kutasov and E. Martinec, 
``On the relevance of tachyons,'' hep-th/0003101.}

The worldsheet analysis (see \hkm\ for a recent discussion) suggests
that there should exist a choice of coordinates on the space of string
fields that is better suited for the study of tachyon condensation. To
see that consider, for example, the process in which the tachyon on
a  $D25$-brane in the bosonic string condenses to make a lower
dimensional $Dp$-brane with $p<25$, which is stretched in the
directions $(1,2,\cdots,p)$. This is achieved by considering the path
integral on the disk with the worldsheet action
\eqn\wstach{\SS=\SS_0+\int_0^{2\pi}{d\theta\over2\pi} T(X(\theta))}
where $\SS_0$ is the free field action describing open plus closed
strings on the disk, $\theta$ is an angular coordinate parametrizing
the boundary of the disk, and $T(X^{p+1}, \cdots, X^{25})$ is a slowly
varying tachyon profile
with a quadratic minimum giving mass to the $25-p$ coordinates
transverse to the $Dp$-brane $\{X^i(\theta)\}$,
$i=p+1,\cdots, 25$. The action
\wstach\ describes a renormalization group flow from a theory where all
$26$ $\{X^\mu\}$ satisfy Neumann boundary conditions (corresponding to the
25-brane) to one where the $25-p$ coordinates $\{X^i\}$ have Dirichlet
boundary conditions (the $p$-brane).

Any profile $T(X)$ with the above properties will do, but a particularly
simple choice is
\eqn\tofx{T(X)=a+\sum_{i=p+1}^{25} u_i X_i^2}
for which the worldsheet theory is free throughout the RG flow.
The parameters $a,u_i$ flow from zero in the UV to infinity in the IR.
A crucial point for what follows is that in this 
flow $a,u_i$  do not mix with any other couplings.

\lref\witbndry{E. Witten, ``On background-independent 
open-string field theory,''hep-th/9208027, Phys. Rev. {\bf D 46} (1992)
5467. ``Some computations in background-independent 
off-shell string theory,'' hep-th/9210065, Phys. Rev. {\bf D 47} (1993) 3405.}
\lref\shat{S. Shatashvili, ``Comment on the background independent 
open string theory,'' hep-th/9303143, Phys. Lett. {\bf B 311} (1993) 83; 
``On the problems with background independence in string theory,''
hep-th/9311177.}

\lref\sencon{A. Sen, ``Descent relations among bosonic 
D-branes,'' hep-th/9902105, Int. J. Mod. Phys. {\bf A14}
(1999) 4061.}

In the spacetime SFT, the $p$-brane can be constructed as
a finite energy soliton \refs{\sencon,\lumps}.
The above worldsheet considerations
imply that there must exist a choice of coordinates on the
space of string fields in which the tachyon profile is given
by \tofx\ and no other fields are excited. In the cubic SFT
\witcs\ this is not the case -- the soliton contains excitations
of an infinite number of fields \sfttach. As we will see below,
a more suitable candidate for describing tachyon condensation
is the open SFT proposed by Witten in \witbndry\ and refined
by Shatashvili \shat\ (see also \liwitten). We will refer to
this string field theory as ``boundary string field theory''
(B-SFT). 

\lref\aflud{I. Affleck and A. W. Ludwig, ``Universal noninteger 
``ground-state degeneracy'' in critical quantum systems,'' \prl{67}{1991}161; 
``Exact conformal field theory results on the multichannel Kondo effect: 
single fermion Green's function, self-energy, and resistivity,'' 
Phys. Rev. {\bf B 48} (1993) 7297.}

The plan of the paper is the following. In section 2 we briefly
review the construction of B-SFT \witbndry. We comment on the
relation of the spacetime action to the boundary entropy of Affleck
and Ludwig \aflud\ and to the cubic Chern-Simons string field theory \witcs.

In section 3 we turn to an example: bosonic open string theory
on a $D25$-brane in flat spacetime. We evaluate the tachyon
potential and kinetic (two derivative) term and study
condensation to the vacuum and to lower branes.
The description of the condensation to the vacuum is exact
(since the tachyon is the only field that condenses, and its
potential is known exactly),  while the  properties of solitons
(corresponding to lower dimensional  branes) receive corrections
from higher derivative terms in the action, although the two
derivative action is in excellent qualitative agreement with
the expected exact results.

In section 4 we show that the corrections to the tension of
solitons in this SFT can be computed exactly, since to analyze
them it is enough to compute the exact action for tachyon profiles
of the form \tofx. We use this observation to compute the tensions
and show that they are in exact agreement with the expected results.
Some comments about the physics of excited open string states and other
issues appear in sections 5, 6. Two appendices contain some of the
technical details.

A. Gerasimov and S. Shatashvili have independently noticed the
relevance of boundary string field theory to the problem of 
tachyon condensation \gerasimov.

\newsec{A brief review of boundary string field theory }

The construction of \witbndry\ is aimed at making precise the
notion that the configuration space of open string field theory
is the space of all two dimensional worldsheet field theories
on the disk, which are conformal in the interior of the disk
but have arbitrary boundary interactions. Thus, as in \wstach,
one studies the worldsheet action
\eqn\sso{\SS=\SS_0+\int_0^{2\pi}{d\theta\over2\pi} \VV}
where $\SS_0$ is a free action defining an open plus closed
conformal background, and $\VV$ is a general boundary perturbation.
We will later discuss the twenty six dimensional bosonic string,
for which $\VV$ has a
derivative expansion (or level expansion) of the form
\eqn\vexp{\VV=T(X)+A_\mu(X)\partial_\theta
X^\mu+B_{\mu\nu}(X)\partial_\theta
X^\mu\partial_\theta X^\nu+C_\mu(X)
\partial^2_\theta X^\mu+\cdots}
The boundary conditions on $X$ (in the unperturbed theory) 
are $\p_r X^\mu\vert_{r=1}=0$. 
If one wishes to include Chan-Paton indices, the field $\VV$ is promoted
to an $N\times N$ matrix and the path integral measure on the
disk is weighted by 
\eqn\measdisk{
e^{-\SS_0} {\rm Tr}{\rm P exp}\biggl( - \int_0^{2\pi}{d\theta\over2\pi}
\VV\biggr)
}
We will mostly restrict to the case $N=1$ in what follows.

In general, $\VV$ is a ghost number zero operator, which nevertheless
might depend on the ghosts, and one must also introduce a ghost number
one operator $\OO$ via
\eqn\vbo{\VV=b_{-1}\OO.}
If, as in \vexp, $\VV$ is constructed out of matter fields alone, one has
\eqn\ocv{\OO=c\VV.}
It is not clear that the theory on the disk described by \sso\ makes
sense. Even if one restricts attention to $\VV$'s that do not depend
on ghosts, such as \vexp, in general the interaction is non-renormalizable
and one might expect the theory to be ill-defined. This is an important
issue,\foot{which is at the heart of the question of background
independence in open SFT, and which was the main motivation for \witbndry.}
about which we will have nothing new to say here, however there
are clearly interesting cases, such as tachyon condensation, in
which the interaction \sso\ is renormalizable. The discussion below
definitely applies to these cases and perhaps more generally.

Parametrizing the space of boundary perturbations $\VV$ by couplings
$\lambda^i$:
\eqn\expv{\VV=\sum_i\lambda^i \VV_i}
(and consequently $\OO=\sum_i\lambda^i \OO_i$ \vbo), the spacetime
SFT action $S$ is defined by
\eqn\sft{ {\partial S\over \partial\lambda^i}={1\over2}
\int_0^{2\pi}{d\theta\over2\pi}\int_0^{2\pi}{d\theta^\prime\over2\pi}
\langle\OO_i(\theta)\{Q,\OO(\theta^\prime)\}\rangle_\lambda}
where $Q$ is the BRST charge and the correlator is evaluated
with the worldsheet action \sso. Note that \sft\ defines the action
up to an additive constant; also, the normalization of the action is
not necessarily the same as in other definitions.\foot{For example,
in the conventional normalization the action goes like $1/g_s$.}
We will fix both ambiguities below.

Specializing to $\OO$'s of the form \ocv, and using the fact that
if $\VV_i$ is a conformal primary of dimension $\Delta_i$,
\eqn\qcomm{\{Q, c\VV_i\}=(1-\Delta_i)c\partial_\theta c\VV_i}
we conclude from \sft\ that
\eqn\dsli{{\partial S\over \partial\lambda^i}=-(1-\Delta_j)\lambda^j
G_{ij}(\lambda)}
where
\eqn\gij{G_{ij}=2
\int_0^{2\pi}{d\theta\over2\pi}\int_0^{2\pi}{d\theta^\prime\over2\pi}
\sin^2({\theta-\theta^\prime\over 2})\langle
\VV_i(\theta) \VV_j(\theta^\prime)
\rangle_\lambda}
Actually, it is clear that eq. \dsli\ cannot be true in general,
since it does not transform covariantly under reparametrizations
of the space of theories, $\lambda^j\to f^j(\lambda^i)$. Indeed, 
$\partial_i S$ and $G_{ij}$ transform as tensors
(the latter is the metric on the space of worldsheet
theories), but $\lambda^i$ does not.
The correct   covariant generalization of \dsli\ was
given in  \shat.
The worldsheet RG defines a natural vector field on the space of
theories
\eqn\rgflow{{d\lambda^i\over d\log|x|}=-\beta^i(\lambda)}
where $|x|$ is a distance scale (\eg\ a UV cutoff), and
\eqn\betfn{\beta^i(\lambda)=-(1-\Delta_i)\lambda^i+O(\lambda^2)}
is the $\beta$ function,\foot{We should note that we are using
the particle physics conventions here -- the $\beta$ function
is negative for relevant perturbations. In some other papers on
the subject, \eg\ \aflud, the opposite conventions are used.}
which transforms as a vector under
reparametrizations of $\lambda^i$.
The covariant form of \dsli\ is thus
\eqn\covs{{\partial S\over \partial\lambda^i}=\beta^jG_{ij}(\lambda).}
As is well known in the general theory of the RG, one can choose
coordinates on the space of theories such that the $\beta$ functions
are exactly linear.\foot{In such coordinates, the BRST charge $Q$ is
independent of the couplings, and \qcomm\ holds everywhere.}
 This can always be done locally in the space of
couplings, so long as the linear term in the
$\beta$-function is non-vanishing. In
such coordinates, \covs\ reduces to \dsli.

In \refs{\witbndry,\shat}
it was further shown that the action $S$ defined by \covs\
is related to the partition sum on the disk $Z(\lambda^i)$ via
\eqn\deftwo{S=(\beta^i{\partial\over\partial\lambda^i}+1) Z(\lambda).}
Note that \deftwo\ fixes the additive ambiguity in $S$ by requiring that
at fixed points of the boundary RG (at which $\beta^i(\lambda^*)=0$)
\eqn\featthree{S(\lambda^*)=Z(\lambda^*).}

{}From the worldsheet point of view, the properties \covs, \deftwo\ and
\featthree\ mean that $S$ is a non-conformal generalization
of the boundary entropy of \aflud. In fact, in any unitary
theory satisfying these properties one can prove the
``$g$-theorem'' postulated in \aflud.
Indeed, the scale variation of $S$ is given by the Callan-Symanzik
equation
\eqn\gthm{{dS\over d\log|x|}=-\beta^i{\partial\over\partial\lambda^i} S
=-\beta^i\beta^j G_{ij}}
where we used the fact that $S$ depends on the scale only via its
dependence on the running couplings, and equations \rgflow, \covs.
In a unitary theory, the metric $G_{ij}$ \gij\ is positive definite;
thus $S$ decreases along RG trajectories. Finally, the property
\featthree\ implies that at fixed points of the boundary RG, $S$
coincides with the boundary entropy as defined in \aflud.
Thus, in any unitary theory in which the considerations of
\witbndry\ do not suffer from UV subtleties (associated with
non-renormalizability), the $g$-theorem of \aflud\ is valid.

As mentioned above, a natural choice of coordinates on the space
of string fields is one in which the $\beta$-functions are exactly
linear. This choice can always be made locally for $\Delta_i\not=1$.
These coordinates become singular as $\Delta_i\to 1$, which in string
theory language is the place where the components of the string field
(\eg\ $T(X)$, $A_\mu$ \etc\ in \vexp) go on-shell. On the other hand,
since the RG flows are straight lines in these coordinates, they are
well suited to studying processes which are far
off-shell, such as tachyon condensation, since they minimize the
mixing between different modes.

In contrast, the cubic SFT parametrization of worldsheet RG is
regular close to the mass shell; it appears to be closely related
to the coordinates on coupling space implied by the
$\epsilon(=1-\Delta)$ expansion.\foot{For a discussion of the
$\epsilon$ expansion in boundary two dimensional QFT see \eg\
\aflud.} These coordinates are useful for studying processes close
to the mass shell, such as reproducing perturbative on-shell amplitudes.

This raises the interesting question of how
 the action $S$ defined above is related to the cubic action
of \witcs. It seems clear that the cubic SFT must correspond to \covs,
\deftwo\
for a particular choice of coordinates on the space of string fields (or
worldsheet couplings).  The  two sets of coordinates are related by
a complicated and highly singular transformation (see appendix A
for some comments on this transformation). As we will see below,
tachyon condensation is simpler in the coordinates \dsli, as one
would expect from the above discussion.

\newsec{A first look at tachyon condensation on the $D25$-brane
in the bosonic string}

In this section we will study the action $S$ described
in the previous section, restricting to the tachyon field.
We will keep terms with up to two derivatives and study various
features of tachyon condensation using the resulting action, which
will turn out to have the form\foot{Our conventions are
$\eta_{\mu\nu}={\rm diag}(-1,+1,\cdots, +1)$.}
\eqn\stact{
S = T_{25} \int d^{26}x \Biggl[ \ap e^{-T} \p_\mu T \p^\mu T
+(T+1) e^{-T} + \cdots \Biggr]
}
where the $\cdots$ stand for terms with more than two derivatives.
Before deriving \stact, we would like to make a few comments on its
form.

\lref\minzwi{J. A. Minahan and B. Zwiebach, 
``Field theory models for tachyon and gauge field 
string dynamics,'' hep-th/0008231.}

\item{(1)} The tachyon potential is
\eqn\tachpot{U(T)=(T+1)e^{-T}.}
This potential is exact, and indeed already 
appears in \refs{\witbndry,\liwitten}.
The perturbative vacuum corresponds
to $T=0$, near which $U(T)=1-{1\over2}T^2+\cdots$.
The ``stable vacuum'' to which the tachyon condenses
is at $T=+\infty$, where $U(T)\to 0$. One can ask why
the tachyon does not instead roll to $T=-\infty$ where
$U(T)$  goes to $-\infty$. We will postpone this issue
to section 6.
\item{(2)} $T_{25}$ in \stact\ is the tension of the
$D25$-brane. Indeed, in the perturbative vacuum $T=0$,
$S=T_{25} V$, where $V$ is the volume of spacetime.
Note that our tachyon field $T$ is dimensionless.
\item{(3)} From \stact\ it seems that the mass of the tachyon in the
perturbative vacuum is $\ap M^2=-1/2$. Of course, the correct result
is $\ap M^2=-1$, but there is no paradox since the higher derivative
terms that have been neglected in \stact\ are important in determining
this mass. In Appendix A we show that the inverse propagator of the
tachyon indeed exhibits a simple pole at $\ap k^2=1$.
\item{(4)} The action \stact\ is related by a field redefinition
\eqn\frbarton{\phi=-2e^{-{T\over2}}}
to an action studied recently in \minzwi\ as a toy model
of tachyon condensation. These authors found that this model
exhibits some remarkable similarities to tachyon condensation
in SFT. We now see that it is in fact a two derivative approximation
to the exact tachyon action. As we will discuss later, this clarifies
the origin of some of the properties found in \minzwi.

\noindent
The action \stact\ can be determined from the definitions \sft, \deftwo\
as follows. One starts by evaluating the partition sum $Z$ in \deftwo\
for the tachyon profile \tofx\ (with generic $u_i> 0$ and $p=-1$).
This should be possible because, as mentioned in section 1, the
resulting worldsheet theory is free for all $a$, $u_i$.
Plugging into \deftwo\ one then finds that the action $S(a, u_i)$
is given by
\eqn\sau{S(a, u_i)=
( -a{\p\over\p a} +1+ \sum 2\ap u_i  - \sum u_i {\p \over \p u_i} )
Z(a, u_i)}
The action \stact\ can then be reconstructed by taking the
limit $u_j\to 0$. A simple scaling argument shows that the
leading behavior of the action \stact\ evaluated on the
profile \tofx\ in the limit $u_j\to 0$ comes from the potential
term; terms with $2n$ derivatives are down from the leading term
by $n$ powers of $u_j$.
Thus, by examining the first two terms in $S$ \sau\ we
can uniquely reconstruct the potential and kinetic terms in \stact.
At higher orders in derivatives, there are many terms one can write down
(most of which vanish on the profile \tofx) and one needs additional
information to determine the full action (see appendix B).

The partition sum $Z(a, u_i)$ has been computed in \witbndry.
The answer can be written as
\eqn\actioni{
S(a,u) = ( a + 1+\sum 2\ap u_i  - \sum u_i {\p \over \p u_i} ) e^{-a}
\prod Z_1(2\ap u_i)
}
where
\eqn\wittenres{
Z_1(u) = \sqrt{u}e^{\gamma u} \Gamma(u).
}
and $\gamma$ is the Euler number.
For small $u_i$ one finds
\eqn\expndactii{
\eqalign{
S & = (a+14) e^{-a} \prod_{j=1}^{26} {1\over \sqrt{2\ap u_j}} \cr
& + 2\ap e^{-a} \bigl( \sum_{j=1}^{26} u_j \bigr) \prod_{j=1}^{26}
{1\over \sqrt{2\ap u_j}} \cr
& + \cdots \cr}
}
The first line of \expndactii\ should be compared to the
potential term in \stact\ evaluated on the profile \tofx;
the second must be due to the kinetic term. Evaluating
the potential energy one finds
\eqn\pe{
T_{25}(2\pi \ap)^{13} e^{-a} (a+14)  \prod_{j=1}^{26} {1\over \sqrt{2\ap
u_j}}
}
Comparing to \expndactii\ we see that
\eqn\teetf{
T_{25} = 1/(2\pi \ap)^{13}.
}
Of course, the fact that this is not the standard form
of the tension is due to the freedom of rescaling
the action \sft, \deftwo. We can use \teetf\ to determine
the multiplicative renormalization of $S$ needed to bring
it into standard form.
Computing the kinetic term in \stact\ and comparing the result
to the second line of \expndactii\ fixes the coefficient
of $(\p_\mu T)^2$ to be the one given in \stact.

After deriving the action \stact\ we are now ready to proceed
to studying tachyon condensation. The first thing that one might
worry about is whether it is enough to study the tachyon dynamics,
or whether one must include the infinite number of excited states,
as in  \sfttach. We will show later that only the zero momentum
tachyon condenses
in the coordinates on string field space that we are working in,
but for now we will assume that and proceed.

Consider first spacetime independent tachyon profiles. The (locally)
stable vacuum is at $T=\infty$. The vacuum energy vanishes there.
Since the potential \tachpot\ is exact,
this gives a proof of Sen's conjecture \sencon\ that
$U_{\rm pert} - U_{\rm closed} = T_{25}$ where  $U_{\rm pert}$ is the
value of the potential in the perturbative open
string vacuum and $U_{\rm closed}$ is the value of the potential
in the closed string vacuum where the open strings have
``condensed'' and ``disappeared.'' We have seen  above that at a
stationary point, $U$  is just the $g$-function, and moreover
that $U_{\rm closed}=0$. On the other hand, it is straightforward
to identify the tension of D-branes with the $g$-function
\refs{\elitzur,\hkms}.

Note that in our coordinates the stable vacuum is at infinity.
This is not in disagreement with other calculations in which it occurs
at a finite value of $T$ \sfttach, since field redefinitions change
the value of $T$ at the minimum of the potential. In fact, in
coordinates on the space of couplings where the $\beta$ functions
in a theory are exactly linear, any infrared fixed points will
{\it always} occur at infinite values of the couplings.

A more invariant question is what is the distance in field space
between the perturbative fixed point at $T=0$, and the stable
minimum at $T=\infty$. For this we need to compute the metric
on the space of $T$'s. This is easily done either by using
\covs\ (with $S=(T+1)\exp(-T)$, $\beta^T=-T$) or by reading off
the metric from the kinetic term in \stact. Either way one finds
that the metric on field space is
\eqn\mett{ds^2=e^{-T} (dT)^2.}
Thus, the distance between $T=0$ and $T=\infty$ is finite.\foot{This
distance is exactly calculable in our approach, and it would
be interesting to compare it to level truncated cubic SFT \sfttach.
This would involve computing the kinetic term for the tachyon and
the other fields that condense in level truncated SFT.}

Consider next spacetime dependent tachyon profiles, which
describe lower dimensional branes as solitons. The equations
of motion following from the action \stact\ are
 \eqn\eomtn{
2\ap \p_\mu\p^\mu T - \ap \p_\mu T \p^{\mu} T + T = 0
}
We are looking for finite action solutions which asymptote to
the ``stable vacuum'' $T=\infty$. The solutions are in fact
precisely the profiles \tofx\ that entered our discussion
a number of times before! Substituting \tofx\ into \eomtn\
one finds that each of the $u_i$ is either $0$ or $1/4\ap$.
That is, the solitons are translationally invariant along
a linear subspace of $R^{26}$ and spherically symmetric
transverse to that subspace.
Let $n$ be the number of nonzero  $u_i$'s. Then $a=-n$.
We interpret such   codimension $n$
solitons as $D(25-n)$-branes.

Substituting into the action \stact\ gives
\eqn\classsol{S=T_{25} (e\sqrt{4  \pi \ap} )^{n } V_{26-n}.
}
Comparing this to the expected tension $T_{25-n} V_{26-n}$ we
conclude that
\eqn\comparten{
{T_{25-n}\over T_{25}} = \biggl(\xi  2 \pi \sqrt{\ap}\biggr)^n
}
with $\xi = e/\sqrt{\pi} \cong 1.534 $. We have written it this
way to facilitate comparison with the exact answer
\eqn\exact{
{T_{25-n}\over T_{25}} = \biggl(  2 \pi \sqrt{\ap}\biggr)^n.
}
In the next section we will see that one can improve on the result
\comparten\ and calculate the tensions \exact\ exactly. But before
moving on to that analysis it is useful to make a few remarks about
the results obtained so far.

One striking feature of the foregoing discussion
is the fact that the soliton solutions are given precisely
by the quadratic tachyon profiles that play such a prominent
role in the worldsheet analysis. This explains why studying them
is so easy: the worldsheet theory in their presence remains free!
It also makes it clear why we are getting descent relations
of the form \comparten: as explained above, the action \stact\
is nothing but the boundary entropy, and for spherically symmetric
profiles of the form
\eqn\solprof{T(X)=-n+u\sum_{i=1}^n X_i^2}
the boundary entropy factorizes. Finally, it is clear why we
are not getting the correct descent relation \exact\ but rather
an approximate version \comparten. The reason is that at this level
of approximation we find a finite value of the mass parameter
$u$ in \solprof. So the action \stact\ is approximately
computing the boundary entropy at a finite point along
the RG trajectory. Since, as discussed in section 2,
the boundary entropy is a monotonically decreasing
quantity, we expect to find a larger answer at finite
$u$ than at the infrared fixed point ($u=\infty$). This
is the reason why the parameter $\xi$ in \comparten\ is
larger than one.

All this makes it clear that what will happen when we include higher
derivative corrections in \stact\ is that the soliton profiles will
still have the form \solprof, but $|a|$ and $u$ will increase to infinity.
We will demonstrate that this is indeed the case in the next section.

The codimension $n$ solitons \solprof\
were also discussed recently in \minzwi. Our results are in
exact agreement with those of \minzwi, although the interpretation
is slightly different. The authors of \minzwi\ analyzed the
spectrum of small fluctuations around the solitons \solprof.
They found a discrete spectrum of scalars with masses
\eqn\masszwi{\ap M_n^2={1\over2}(n-1);\;\;n=0,1,2,\cdots}
This is very natural from the worldsheet point of view as well,
since once we turn on a worldsheet potential of the form \solprof\
(even for finite $u$),
it is clear that one expects to find only fields that are bound to the
lower dimensional brane but otherwise have the same properties
as their higher dimensional cousins.

Finally, one might wonder whether it is possible to describe
multi-soliton configurations in the theory \stact. From the
worldsheet point of view this involves \hkm\ studying multicritical
tachyon profiles of the form
\eqn\multicrit{T(X)=\sum_{j=1}^la_j|\vec X|^{2j}.}
For $l>1$ the worldsheet theory is no longer free and one expects
complications having to do with the interactions between
the solitons (fundamental strings connecting different
D-branes).
Plugging \multicrit\ into \eomtn\ we see that the reflection
of this in spacetime is that one needs to keep higher derivative
terms in the action to study such configurations.

\newsec{An exact calculation of  D-brane tensions  }

We would like to compute the corrections to the descent relations
\comparten\ coming from higher derivative corrections to the
action \stact. In principle, one might proceed as follows. First
generalize the procedure of section 3 to compute higher derivative
corrections to the action, and then use the resulting action
to determine the profile of the solitons and their tensions.
This looks difficult; computing the higher derivative corrections
involves both technical and conceptual complications. Also it is likely
that the resulting action would be rather unwieldy and difficult
to study.

Some of the technical complications can be seen by looking at
the action for quadratic tachyon profiles \actioni, \wittenres.
As discussed in section 3, implicit in the action $S(a, u_i)$
is an infinite series of higher derivative corrections to
\stact\ which can be computed by expanding $S$ in powers of
the $u_j$. An example of such an infinite-derivative 
action is given in appendix B. 

Unfortunately,  \actioni\ and \wittenres\
do not determine the tachyon action uniquely, since it is easy to
write an infinite number of terms which annihilate the profile
\tofx\ and thus do not contribute to $S(a, u_i)$.
Nevertheless, the discussion of the previous section makes it
clear that there is an alternative way to proceed that
circumvents all of the above complications and can be used
to compute the tensions of the solitons exactly. The basic
observation is that {\it we know that the exact profile of the
soliton in the exact SFT \covs, \deftwo\ is going to be
of the form \tofx, with some particular values of $a$, $u_i$}.
The reason is that this mode does not mix with any other modes
in the SFT (as will be shown in the next section).

Thus, all we have to do to compute the exact tension of the
D-brane solitons is to take the exact action $S(a, u_i)$
given by \actioni, \wittenres, and extremize it in $a$ and $u_i$.
Furthermore, we know that the extremum we are looking for
is one in which $n$ of the $u_i\to \infty$ and the rest
vanish (for a codimension $n$ soliton). We next describe
this calculation.

For simplicity let us consider first a codimension one
soliton. We would like to substitute the ansatz
$T= a + u X_1^2$ in \stact\ (with the other $u_i=0$)
and set the action equal to $V_{24+1} T_{24}$. Of course,
the action \actioni, \wittenres\ diverges when $u_i\to 0$,
which  is a reflection of the divergent volume
$V_{24+1}$. In order to do the computation in a well defined
way we must regularize the volume divergence. We do this by
periodic identification of
\eqn\periodic{
X^\mu \sim X^\mu + R^\mu \qquad\qquad  \mu = 2, \dots, 26.
}
We must now determine the correct normalization of the
path integral $Z$. The correct normalization for the
worldsheet
zero-mode of an uncompactified  spacetime coordinate $X$ is
\eqn\normzero{
 \int_{-\infty}^{\infty}{d X \over \sqrt{2\pi\ap}}
e^{-\int_{0}^{2\pi} {d\theta\over 2\pi} T(X(\theta))}.
}
We know this because if we substitute $T=a + u X^2 $,
we reproduce $e^{-a} {1\over \sqrt{2\ap u}}$.
It follows that when we periodically
identify $X^\mu$ as in \periodic\ in directions $\mu=2,\dots, 26$ and
take $T= a+ u X_1^2$ the resulting boundary string field theory
action is, exactly,
\eqn\bdrysft{
S = \Biggl( a+ 1 - u {\p \over \p u} + 2\ap u\Biggr)
e^{-a} Z_1(2\ap u) \prod_{\mu=2}^{26} \biggl({R^\mu \over \sqrt{2\pi
\ap}}\biggr).
}
As discussed above, the dynamical variables in this action
are $a,u$. Therefore, we should minimize $S$ with respect to
them. Minimizing first with respect to $a$ we find
\eqn\ainyou{
a_* = a(u) = - 2\ap u + u{d \over d u} \log Z_1(2\ap u).
}
Substituting back into the action we get:
\eqn\backinto{
S = \exp[ \Xi(2\ap u)  ] \prod_{\mu=2}^{26} \biggl({R^\mu \over \sqrt{2\pi
\ap}}\biggr)
}
where we define
\eqn\defsxiu{
\Xi(z) :=  z - z{d\over dz} \log Z_1(z) + \log Z_1(z).
}
We may now invoke Witten's result \wittenres.
The action \backinto\ is a monotonically decreasing function of $u$,
and therefore the minimization pushes $u$ to $\infty$, as expected
from the worldsheet renormalization group arguments (the $g$-theorem).

We are particularly interested in the value of the action
at the end of the RG trajectory.
{}From Stirling's formula we find at large $z$
\eqn\stirling{
\eqalign{
\log Z_1(z) & \sim z \log z -z + \gamma z + \log \sqrt{2\pi} + \CO(1/z),\cr
\Xi(z) & \sim  \log \sqrt{2\pi}  +1/(6z) + \CO(1/z^2). \cr}
}
We thus obtain the boundary string field theory action
\eqn\valuebdry{
\sqrt{2\pi} \prod_{\mu=2}^{26} \biggl({R^\mu \over \sqrt{2\pi \ap}}\biggr)
}
On the other hand, from the spacetime point of view this is clearly
equal to $T_{24}\prod_\mu R^{\mu}$. We therefore conclude
that
\eqn\finaltension{
T_{24} = 2 \pi \sqrt{\ap} T_{25}
}
which is {\it precisely} the expected value!

Clearly this exercise can be repeated for branes of higher codimension.
After minimization with respect to $a$ we find the action for the
codimension $n$ soliton:
\eqn\bdrcodenn{
\exp\bigl[\sum_{i=1}^n \Xi(2\ap u_i) \bigr] \prod_{\mu=n+1}^{26}
\biggl({R^\mu \over \sqrt{2\pi \ap}}\biggr)
}
and therefore each codimension leads to an extra factor of $2 \pi
\sqrt{\ap}$, in agreement with \exact.

\noindent
We finish this section with a few comments:

\item{(1)} The solitonic solutions describing lower dimensional
D-branes constructed in section 3 had a finite size, of order
$l_s=\sqrt{\ap}$ (since their profiles were given by \solprof\
with $u=1/4\ap$). In the exact problem, the sizes of the solitons
go to zero like $1/\sqrt u$. This is in nice correspondence with the
usual description of D-branes as (classically) pointlike objects.
In level truncated SFT, the lower D-branes were found to correspond
to finite size lumps, similar to those of section 3. Here, we saw
that the higher derivative terms in the action play a crucial role
in reducing the size of the soliton from $l_s$ to zero. Since in
the level truncation scheme the contributions of such terms 
seem to increase with level, it is possible that if the calculations
of \lumps\ were continued to much higher levels, the size of the
solitons would slowly decrease to zero, as it does in our approach.
Another possibility is that the complicated relation of our
parametrization of the space of string fields to that of cubic
SFT transforms the $\delta$-function tachyon profile we find to
a finite size lump.

\item{(2)} The fact that we have been able to  reproduce exactly
the tension ratios \exact\ may at first sight seem puzzling.
The full spacetime classical SFT is a very complicated
theory, with an infinite number of fields and a rich pattern
of non-polynomial interactions. The fact that one can prove that
this theory has finite action solitonic solutions with profiles
and tensions that can be computed exactly looks from the spacetime
point of view like a ``string miracle.''
Such ``miracles'' are very
generic in string theory. The oldest example is perhaps
(channel) duality of the tree level S-matrix.
The fact that an infinite sum over massive s-channel poles
can produce a t-channel pole is due in spacetime to an incredible
conspiracy of the masses and couplings of Regge resonances.
Describing this in terms of a spacetime
Lagrangian seems hopeless. However,
on the worldsheet, this is one of the many consequences of
conformal invariance and is easily described and understood.
In the tachyon condensation problem, something very similar
happens. The miracle is explained by noting that the spacetime
action is nothing but the boundary entropy (see section 2),
and the process of condensation is trivial since it corresponds
to free field theory on the worldsheet \tofx.

\newsec{Comments on excited open strings}

In our discussion so far we focused on the physics of
the tachyon. It is interesting, and for some purposes
necessary, to generalize the discussion to include excited
open strings.

The first question that we address is one
that was noted a few times in the text: why can we study
condensation of the tachyon without taking into account
other modes of the string? The reason is that we can 
divide the coordinates on field space into $a,u$, 
which are free field perturbations and an orthogonal 
set of coordinates $\lambda_i$ corresponding to the 
non-zero momentum modes of the tachyon and excited
open string modes. The $\lambda_i$ could be \eg\ modes
of one of the fields $A_\mu$, $B_{\mu\nu}$, $C_\mu$ in \vexp. 
It is consistent to  set all the
excited string modes $\lambda_i$ to zero in the presence of a tachyon
profile of the form  \tofx\ if and only if the action
\covs, \deftwo\ does not have any linear terms in any of the couplings
$\lambda_i$ in the background \tofx.  

It should be emphasized that while the couplings $\lambda_i$  
are in general non-renormalizable (since they correspond to
irrelevant operators with $\Delta_i>1$), we are treating the
dependence of $S$ on $\lambda_i$ perturbatively. There is no
problem with calculating integrated correlation functions of
irrelevant operators in a background such as \sso, perturbed
from a conformal background by relevant and marginal operators,
at least after suitable regularization and renormalization
procedures are specified. One such procedure is described in
appendix A (by contrast, studying a worldsheet action
like \sso\ with a finite perturbation by 
an irrelevant operator is likely to lead to inconsistencies.)
Accordingly, we may write the action in the form
\eqn\saull{S(a, u, \lambda^i)=S^{(0)}(a, u)+
S^{(1)}_i(a, u)\lambda^i+S^{(2)}_{ij}(a,u) \lambda^i \lambda^j +
\cdots
}
where $S^{(0)}(a, u)$ is the action \actioni\ and we would like
to prove that $S^{(1)}_i(a, u)=0$. Suppose, on the contrary
that $S^{(1)}\not=0$.
Then $\partial_i S|_{\lambda=0}\not=0$. Looking back at
equation \covs\ we see that this means that if the metric $G_{ij}$
is non-degenerate on the space of couplings orthogonal 
to $a,u$, then $\beta^j(a,u; \lambda^i=0)\not=0$.

Now, after fixing string gauge invariances, 
the metric $G_{ij}$ is   non-degenerate in
the background \tofx, which corresponds to
free field theory. At the same time, the statement
that $\beta^i(\lambda)$ does not vanish at $\lambda^i=0$
implies that as we turn on $a$ and the $u$'s, the
$\lambda^i$ start flowing according to \rgflow. But
we know that this is false. In free field theory no
new couplings are generated by the RG flow.
Therefore, $S^{(1)}_i(a,u)$ must vanish. 
We conclude that all other string modes appear at least
quadratically in the spacetime action in the tachyon
backgrounds \tofx, and they can be consistently set to
zero when studying tachyon condensation.

Again, it is interesting to contrast the situation with the 
cubic SFT. In this case a higher string mode, call it schematically 
$v$, can couple to the tachyon $T$ schematically 
as $v^2 + v T^2 + v^2 T + v^3$. The couplings of the form 
$v T^2$ are generically nonzero, and indeed the explicit computations 
\refs{\sfttach,\lumps} show that higher string modes do obtain nontrivial
expectation values during tachyon condensation. 

\lref\sssee{A. Sen, ``Supersymmetric world-volume action for 
non-BPS D-branes,'' hep-th/9909062, JHEP {\bf 9910} (1999) 008. }

Another interesting circle of questions surrounds the fate of the
excited string modes as $T\to \infty$. From the worldsheet analysis
it is expected that they all ``disappear'' from the spectrum,
but the precise mechanism by which this happens in spacetime
is not well understood. It has been proposed \sssee\ that the
coefficients of the kinetic terms vanish at the ``stable minimum''
but the situation is unclear. The viewpoint of this paper sheds 
some light on these issues. 

We would like to construct the action
for excited open string modes using the 
prescription \covs, \deftwo.
We may determine the dependence on the
zero mode of the tachyon as follows.
Consider the theory in the
background $T=a$ (corresponding to \tofx\ with $u_i=0$).
The partition sum has in this case a simple dependence on $a$,
\eqn\zalambda{Z(a, \lambda_i)=e^{-a} \tilde Z(\lambda_i)}
where we denoted all the other modes collectively by
$\lambda_i$.
The action \deftwo\ therefore takes the form
\eqn\salambda{S(a, \lambda_i)=(a+1+\beta^i{\p\over\p\lambda_i})
e^{-a}\tilde Z(\lambda_i).}
Recalling the form of the exact tachyon
potential \tachpot\ the action \salambda\ can be rewritten
as
\eqn\ssaall{S(a,\lambda_i)=U(T)\tilde Z(\lambda_i)+
e^{-T}\beta^i{\p\over\p\lambda_i}\tilde Z(\lambda_i)}
As we show in appendix A, near the mass shell (\ie\
as $\Delta_i\to 1$), the quadratic term in the partition
sum exhibits a first order zero $(\propto 1-\Delta_i)$; 
thus the usual kinetic terms
for the modes $\lambda_i$ come from the first term on the
r.h.s., while the second term, which goes like $(1-\Delta_i)^2$
near the mass shell, contributes higher derivative corrections.
In any case, we see that all terms in the action go to zero
as the tachyon relaxes to $T=\infty$, but, at least in these
coordinates on the space of string fields, they do not all go like
$U(T)$.

\lref\tseyt{A. Tseytlin, ``Born-Infeld action, supersymmetry 
and string theory,'' hep-th/9908105.}

A simple application of \ssaall\ is to the dependence of the
Born-Infeld action on $T$ discussed in \sssee. A constant
$F_{\mu\nu}$ on the $D25$-brane does not break conformal
invariance, and therefore the second term in \ssaall\
vanishes in this case. The partition sum in
the presence of the constant electro-magnetic field
is the Born-Infeld action (for a review see \tseyt),
\eqn\zbi{Z(F)=\LL_{BI}(F).}
Substituting into \ssaall\ we conclude that
the action for slowly varying gauge fields and tachyons
is
\eqn\slvar{S=U(T)\LL_{BI}}
in agreement with \sssee\ (essentially the same result
already appears in \liwitten.) 

One can also use our construction to study the spectrum
of the open string theory in the background of a soliton.
This involves computing the partition sum $Z$ to quadratic
order in the couplings $\lambda_i$ in the soliton background
\tofx\ and should give rise to the standard picture of states
bound to the soliton (or lower dimensional brane).
As we have mentioned, this should help to explain some 
results of \minzwi. 

\newsec{Many open problems}

There is a large number of open problems associated with the
circle of ideas explored in this paper. In this section we
list a few.

It would be interesting to calculate additional
terms in the SFT action. This involves both the determination
of higher derivative corrections to \stact\ and the inclusion
of excited string modes discussed in the previous section.
As noted in the text, the exact action \actioni\ implies
an infinite number of higher derivative corrections to
\stact, but in order to
calculate all terms of a given order in derivatives,
more information is needed. Perhaps, additional information
can be obtained by solving the worldsheet theory corresponding
to the multi-soliton tachyon profiles \multicrit.

A related problem is understanding more clearly the relation 
between boundary string field theory  and the cubic SFT. 
It is conceivable that the space of 2d field theories is a 
nontrivial infinite dimensional space with no good global 
coordinate system. It appears from the singular relation 
between the fields (see \eg\ appendix A) that coordinates
appropriate to the cubic 
SFT might have a range of validity which is geodesically 
incomplete and does not coincide with the ``patch'' in 
which good coordinates for boundary SFT are valid. 

The discussion throughout this paper has focused on the bosonic
string, but the construction of section 2 is more general. In
particular, the worldsheet RG picture has been generalized to
the superstring \hkm, where it applies to non-BPS $D$-branes,
$D-\bar D$ systems and related configurations. It would be
interesting to generalize the considerations of this paper to
these problems, especially because the generalization of the
cubic SFT to the superstring is subtle and complicated.

Another interesting problem involves the role of quantum effects
in the tachyon condensation process.\foot{We thanks T. Banks and
S. Shenker for a discussion of this issue.} Our discussion here was
entirely classical, and yet we found that the action goes to
zero as the tachyon condenses \ssaall. Usually this is taken
to be a sign of strong coupling, and indeed there were proposals
in the literature that tachyon condensation leads to a strongly
coupled string theory. For example, the form \slvar\ of the gauge
field action seems naively to suggest that the effective Yang-Mills
coupling behaves as
\eqn\effgym{{1\over g^2_{YM}}={U(T)\over g_s}}
and therefore, as $U(T)\to 0$ the gauge theory becomes
more and more strongly coupled.

On the other hand, the worldsheet analysis of \hkm\ seems
to suggest that no strong coupling behavior should
be encountered as the tachyon condenses, since diagrams
with many holes are not becoming larger in this process.

Boundary SFT seems to lead to the same conclusion. It is 
natural to expect that quantum
corrections to the string field action $S$ \deftwo\ come
from performing the worldsheet path integral over Riemann
surfaces with holes. Each hole contributes a factor of $g_sN$
as usual (for $N$ $D25$-branes), as well as a factor of
$\exp(-T)$ from the path integral of \wstach.
Thus, it looks like the effective coupling is in fact
\eqn\effcoup{\lambda=g_sNe^{-T}}
and the perturbative expansion looks like
\eqn\perturb{
Z = 
N^2e^{-2 T} \biggl( {1\over \lambda} A_{-1} +
A_0 + \lambda A_{1} + \cdots\biggr)
}
where $A_{-\chi}$ is obtained
from the path integral on surfaces of Euler character $\chi$
and no handles. Eq. \perturb\ suggests that the theory
in fact remains weakly coupled as $T\to\infty$, but this seems
difficult to reconcile with the Feynman diagram expansion
arising from the coupling \effgym.
It would be interesting to resolve this apparent contradiction.

The behavior of the effective string coupling 
\effcoup\ is related to another
issue raised earlier in the paper.\foot{We thank P. Horava
and H. Liu for useful comments on this issue.}
Recall that the tachyon
potential  \tachpot\ is not bounded from below as $T\to-\infty$.
Even if the tachyon condenses to the locally stable
vacuum at $T=+\infty$ (the closed string vacuum), the system
will tunnel through the potential barrier to the true
vacuum at $T=-\infty$. The instanton responsible for
this tunneling is the Euclidean bounce solution 
corresponding to a codimension twenty six brane in our
construction. Like all the other solitons, it has one negative
mode, and therefore mediates vacuum decay. It is natural to 
ask what is the nature of this instability. 

The behavior of the effective coupling \effcoup\ provides a hint
for a possible answer. We see that as the
system rolls towards the ``true vacuum'' at $T\to-\infty$, the
string coupling grows. This is significant since
as is well known, quantum mechanically, open strings can
produce closed strings, and in particular, in this case, 
the closed string tachyon. Thus, one is led to interpret the
instability of the ``closed string vacuum'' at $T\to\infty$
to decay to the ``true vacuum'' at $T\to-\infty$ as the closed
string tachyon instability. While this is a speculation that
needs to  be substantiated, we note the following as (weak) evidence
for it:
\item{(1)} The amplitude for false vacuum decay due to the
bounce goes like $\exp(-1/g_s)$. The fact that it vanishes
to all orders in $g_s$ is consistent with the fact that no
such instability is observed in perturbative open string theory
\hkm. Understanding the precise dependence on $g_s$ probably
requires a better understanding of the issues discussed around
equation \effcoup.
\item{(2)} The fact that the string coupling grows  
after closed string tachyon condensation, suggested by \effcoup,
is consistent with the known physics of closed string tachyon
condensation. In this process the central charge of the system
decreases, and the dilaton becomes non-trivial (linear in one
of the coordinates). This leads to strong coupling somewhere in
space.
\item{(3)} For unstable $D$-branes in the superstring, the
corresponding tachyon potential does not have a similar
instability, in accord with the fact that there is no
tachyon in the closed string sector in that case.

\bigskip
\noindent{\bf Acknowledgements:}
We would like to thank T. Banks, M. Douglas, J. Harvey,
E. Martinec, N. Seiberg, S. Shenker, and A. Strominger
for useful discussions, and J. Harvey for comments on
the manuscript. We also thank the participants of
the Rutgers group meeting for many lively questions during
a presentation of these results. The work of D.K. is
supported in part by DOE grant \#DE-FG02-90ER40560. The work
of GM and MM is supported by DOE grant DE-FG02-96ER40949.
DK thanks the Rutgers High Energy Theory group for hospitality
during the course of this work. MM would like to thank 
the High Energy Theory group at Harvard for hospitality
in the final stages of this work.

\appendix{A}{Some technical results}

The string field theory 
action \sft\ for a tachyon profile 
\eqn\taprof{
T(X)=\int dk \, \phi(k)\, \exp (ik\cdot X)}
is an infinite series $S^{(2)} + S^{(3)} + S^{(4)} + \cdots $ 
in powers of $T$. In this appendix we give explicit formulae 
for the first two terms in this expansion.  
More generally, we will show that the quadratic term in the 
string field action for a primary field $\VV_i$ has a pole
at $\Delta_i=1$.  

The structure of the quadratic term $S^{(2)}$ for a primary 
field $\VV_i$ has a rather simple expression. We only need the
correlation function of the boundary operator 
in the free field theory, which is given by 
\eqn\corrprim{
\langle \VV_i(\theta)\VV_i(\theta') \rangle = {c \over
\biggl[ \sin^2 \bigl( {\theta -\theta' 
\over 2} \bigr) \biggr]^{\Delta_i}},} 
and $c$ is a constant. We also need the value of the 
following integral: 
\eqn\inte{
\int_0^{2 \pi} {d \theta\over 2\pi} \int_0^{2\pi} {d \theta' \over 2\pi} 
\biggl[ \sin^2 \bigl( {\theta -\theta' 
\over 2} \bigr) \biggr]^z = 4^{-z}{\Gamma(1+2z) \over \Gamma^2(1+z)} 
={1\over \pi^{1/2}}{\Gamma(1/2+z) \over\Gamma(1+z)}.} 
Notice that 
in the evaluation of this integral we have regulated the 
short-distance singularities by analytic continuation in $z$. 
We will now compute $S^{(2)}$ using 
\deftwo. The term of order $(\lambda^i)^2$ 
in the partition function is given by:
\eqn\ztwo{
Z^{(2)}={ c \over 2} (\lambda^i)^2 \int_0^{2 \pi} {d \theta\over 2\pi} \int_0^{2\pi} {d \theta' \over 2\pi} 
\biggl[ \sin^2 \bigl( {\theta -\theta' 
\over 2} \bigr) \biggr]^{-\Delta_i} = 
{c (\lambda^i)^2 \over \pi^{1/2}}{ \Gamma(3/2-\Delta_i) 
\over (1-2\Delta_i) \Gamma(1-\Delta_i)}.}
We see that this gives a simple pole in the propagator at
$\Delta_i=1$, a fact that was used in section 5. The action
$S$ \deftwo\ to quadratic order is then:
\eqn\stwo{
S^{(2)}=(1-2(1-\Delta_i))Z^{(2)}=-{c (\lambda^i)^2 \over 
\pi^{1/2}}{ \Gamma(3/2-\Delta_i) 
\over  \Gamma(1-\Delta_i)}.} 
Notice that the term $\beta_i \partial^i Z$ gives a second order pole 
in the propagator at this order, as stated in section 5. It is easy to check 
that the definition \covs\ gives the same answer for $S^{(2)}$.  
 
In the case of the tachyon field \taprof, the correlation function 
in free field theory is  
\eqn\correl{
\langle {\rm e}^{i k\cdot X(\theta)} {\rm e}^{i k'\cdot X(\theta')} 
\rangle = (2\pi)^d \delta(k+k') \biggl[ \sin^2 \bigl( {\theta -\theta' 
\over 2} \bigr) \biggr]^{-\alpha' k^2},}
and the quadratic piece of the action \stwo\ reads in this case
\eqn\quad{
S^{(2)}=-\int dk \, {1 \over \pi^{1/2}} {\Gamma (3/2 -\alpha' k^2) 
\over  \Gamma(1-\alpha' k^2)}(2 \pi)^d \phi(k) \phi(-k).}
Notice again that the propagator, which is a complicated function 
of $k^2$, exhibits the required pole at $\alpha'k^2 =1$.

The cubic term for the tachyon field \taprof\ can be 
computed by evaluating the correlation function \correl\ at 
next order in perturbation theory. The result is:
\eqn\third{\eqalign{
S^{(3)}=& -{1 \over 3!}
\int dk d k ' d k'' \,  
\phi(k)\phi(k') \phi(k'')(2\pi)^d\delta(k+k'+k'') 
4^{-(a+b+c)+1}(1-\alpha'k'^2)\cr
&\cdot {\Gamma(1+2a)\Gamma(1+2b)
\over \Gamma^2(1+a-c)\Gamma^2(1+b-c)} 
{}_{3}F_{2} (-2c, -a-c, -c-b; 1+a-c,1+b-c;1),\cr}}
where 
\eqn\coefs{
\eqalign{
a=&-\alpha'k\cdot k'+1,\cr
b=&- \alpha'k\cdot k'',\cr
c=&-\alpha'k'\cdot k'',\cr}}
and the hypergeometric function ${}_pF_{q}$ is 
defined by 
\eqn\defh{
{}_pF_{q}(\alpha_1, \cdots, \alpha_p;\beta_1,\cdots, \beta_q;z)
=\sum_{n=0}^\infty  \prod_{i=1}^p\biggl({\Gamma(\alpha_i+n)\over 
\Gamma (\alpha_i)}\biggr)\prod_{j=1}^q\biggl({\Gamma(\beta_j)\over 
\Gamma (\beta_j+n)}\biggr){z^n\over n!}.} 

We can now try to compare the action $S=S^{(2)} + S^{(3)}+ \cdots$ 
to the cubic action obtained 
in the open string field theory of \witcs. Using, for 
example, the approach of \lcpp, 
one finds 
\eqn\cubic{
\eqalign{
S^{CS}&=A \int dk \, (2\pi)^d {\widehat \phi}(k) {\widehat
\phi}(-k)(\alpha'k^2 -1) \cr  
&+\int dk d k ' d k'' \,  B(k,k',k'')
{\widehat \phi}(k){\widehat \phi}(k') {\widehat \phi}(k'')(2\pi)^d 
\delta(k+k'+k''),\cr}}
where  
\eqn\coeefs{
A=-{1 \over 2 g_s^2}, \,\,\  B(k,k',k'')=-{1 \over 3 g_s^2} 
\biggl( {4 \over 3 {\sqrt 3}} \biggr)^{-a-b-c-2}.}
If we assume that the tachyon fields $\phi(k)$ and ${\widehat \phi}(k)$ 
are related as follows,
\eqn\rela{
  {\widehat \phi}(k)=f_1(k) \phi(k) + \int dk' f_2(k,k') 
\phi(k') \phi(k-k') + \cdots,}
in such a way that 
\eqn\relatwo{
S^{CS}({\widehat \phi}(k))= \kappa \, S(\phi(k)),}
where $\kappa$ is a nonzero constant, we obtain:
\eqn\relathree{
(f_1(k^2))^2= {\kappa \over \pi^{1/2} A} {\Gamma(3/2 -\alpha'k^2) 
\over \Gamma (2 -\alpha'k^2)},}
where we have used that $f_1(k) =f_1(-k)$ (this follows from 
reality of the tachyon field). By comparing the cubic terms, 
we find:

\eqn\compl{
\eqalign{
&A(\alpha'k^2 -1)(f_1(k)f_2(-k,k'')+ f_1(k')f_2(-k',k'')) + 
B(k,k',k'')f_1(k) f_1(k')f_1(k'')\cr
&={\kappa \over 3!}4^{-(a+b+c)+1}(\alpha'(k')^2 -1)G(k,k',k''),\cr}}
where we have defined
\eqn\func{
G(k,k',k'')={\Gamma(1+2a)\Gamma(1+2b)
\over \Gamma^2(1+a-c)\Gamma^2(1+b-c)} 
{}_{3}F_{2} (-2c, -a-c, -c-b; 1+a-c,1+b-c;1).}
Notice that $f_1(k)$ is regular and different from zero 
when the tachyon is on-shell. On the other hand, if we evaluate 
the relation \compl\ when the three tachyon fields are on-shell,  
we find that $G(k,k',k'')=0$, and therefore $f_2(k,k'')$ must 
have a pole with nonzero residue at $\alpha'k^2=1$. This shows 
that the relation between the CS and the B-SFT tachyon fields
becomes singular on-shell.

\appendix{B}{Some higher derivative terms in the tachyon action}

In this appendix we give an example of a higher derivative 
Lagrangian for the tachyon which reproduces the exact 
action $S(a,u)$. This is simply meant to indicate the nature 
of some of the terms. We stress at the outset that the 
following does not determine an infinite set of couplings, 
namely, anything which vanishes on the Gaussian profile. 
One unambiguous conclusion one can draw from this exercise 
is that in terms of $\phi\sim e^{-T}$ the higher derivative 
terms must be singular at $\phi=0$. 

It is useful to generalize the tachyon profile to 
$T= a + u_{\mu\nu} X^{\mu} X^{\nu}$ with $u_{\mu\nu}$ 
positive definite. The exact action may be written as 
\eqn\hdi{
S = \bigl[ a + 1+  2\ap {\Tr}(u)  - {\Tr}(u {\p \over \p u})\bigr]  
{e^{-a} \over \sqrt{\det u}}  \exp \biggl[ 
\sum_{k=2}^{\infty} {(-1)^k \zeta(k) \over k} {\Tr u^k} \biggr]}
(there is a regularization dependent term $\sim {\rm Tr} u$
in the exponential. With the normal ordering prescription of
\witbndry\ this term vanishes).
Expanding the exponential we obtain a series 
\eqn\hdii{ 
 \sum_{n_k\geq 0 } A_{n_k} \prod_k ({\Tr}u^k)^{n_k}= 
1+ \half \zeta(2) {\Tr}(u^2) + \cdots 
}
where at a given order in scaling under $u \to \alpha u$ 
the sum is a Schur polynomial.

Now the action becomes: 
\eqn\hdiii{
\eqalign{
a e^{-a} {1\over \sqrt{\det u}} &  \sum_{n_k} A_{n_k} \prod_k ({\Tr}u^k)^{n_k} \cr
+e^{-a} 
{1\over \sqrt{\det u}} &  \sum_{n_k} A_{n_k}(14 -L_0(n_k)) \prod_k ({\Tr}u^k)^{n_k} \cr
+2\ap ({\Tr} u) 
e^{-a} {1\over \sqrt{\det u}} &  \sum_{n_k} A_{n_k} \prod_k ({\Tr}u^k)^{n_k} \cr}
}
where it is convenient to define $L_0(n_k) = \sum_k k n_k$. 

One straightforward 
way to reproduce this from a Lagrangian proceeds by 
starting with the $a e^{-a}$ term.  This is reproduced by 
\eqn\hdiv{
T_{25}\pi^{-13} \int dx~~ T e^{-T} \sum_{n_k} 
\bigl({1\over 2}\bigr)^{L_0(n_k)} 
A_{n_k} \prod_k \bigl[(\p_{\mu_1\mu_2} T)(\p_{\mu_2\mu_3} T) \cdots (\p_{\mu_k \mu_1} T) \bigr]^{n_k}
}
In order to account for the second line in \hdiii\ we  
 add terms 
\eqn\hdv{
T_{25}\pi^{-13} \int dx
e^{-T} \sum_{n_k} \bigl({1\over 2}\bigr)^{L_0(n_k)} 
B_{n_k} \prod_k \bigl[(\p_{\mu_1\mu_2} T)(\p_{\mu_2\mu_3} T) \cdots (\p_{\mu_k \mu_1} T) \bigr]^{n_k}
}
with   $B_{n_k} = (1-L_0(n_k))A_{n_k} $. 
Finally to get the last line of \hdiii\ we take
\eqn\hdvi{
\ap T_{25}\pi^{-13} \int dx~~
e^{-T} (\p_{\mu} T)(\p_{\mu} T)
\sum_{n_k} \bigl({1\over 2}\bigr)^{L_0(n_k)} 
A_{n_k} \prod_k \bigl[(\p_{\mu_1\mu_2} T)(\p_{\mu_2\mu_3} T) \cdots (\p_{\mu_k \mu_1} T) 
\bigr]^{n_k}
}

\listrefs
\bye